\documentclass[letterpaper, 10 pt, conference]{ieeeconf}  

\IEEEoverridecommandlockouts                              
\usepackage{amsmath}
\usepackage{amssymb}
\usepackage{mathtools}
\usepackage[dvipsnames]{xcolor}
\usepackage{tikz}
\usepackage[implicit=false]{hyperref}

\newtheorem{thm}{Theorem}
\newtheorem{lem}{Lemma}

\newtheorem{defn}{Definition}
\newtheorem{rem}{Remark}
\newtheorem{assum}{Assumption}
\newtheorem{prob}{Problem}

\newcommand{\R}{\mathbb{R}}
\newcommand{\mL}{\mathcal{L}}
\newcommand{\mH}{\mathcal{H}}

\DeclareMathOperator*{\diag}{diag}
\DeclareMathOperator*{\blkdiag}{blkdiag}
\DeclareMathOperator*{\rank}{rank}

\usetikzlibrary{shapes,arrows,calc,positioning}
\tikzset{
input/.style={coordinate},
midpoint/.style={coordinate},
arrow/.style={draw,-latex,thick},
arrow2/.style={draw,latex-latex,thick},
line/.style={draw,-,thick},
sum/.style={draw,circle,node distance=1cm,thick}
}

\newcommand\copyrighttext{%
	\footnotesize \copyright 2024 IEEE. Personal use of this material is permitted. Permission from IEEE must be obtained for all other uses, in any current or future media, including reprinting/republishing this material for advertising or promotional purposes, creating new collective works, for resale or redistribution to servers or lists, or reuse of any copyrighted component of this work in other works.}
\newcommand\copyrightnotice{%
	\begin{tikzpicture}[remember picture,overlay]
		\node[anchor=south,yshift=10pt] at (current page.south) {\fbox{\parbox{\dimexpr\textwidth-\fboxsep-\fboxrule\relax}{\copyrighttext}}};
	\end{tikzpicture}%
}

\title{\LARGE \bf
Data-Based System Representation and Synchronization \\ for Multiagent Systems
}

\author{Victor G. Lopez and Matthias A. Müller
\thanks{This work received funding from the European Research Council (ERC) under the European Union’s Horizon 2020 research and innovation programme (grant agreement No 948679).}
\thanks{V. G. Lopez and M. A. Müller are with the Leibniz University Hannover, Institute of Automatic Control, 30167 Hannover, Germany
        {\tt\small \{lopez,mueller\}@irt.uni-hannover.de}}%
}

\begin{document}

\maketitle
\thispagestyle{empty}
\pagestyle{empty}
\copyrightnotice

\begin{abstract}

This paper presents novel solutions of the data-based synchronization problem for continuous-time multiagent systems. We consider the cases of homogeneous and heterogeneous systems. First, we obtain a data-based representation of the synchronization error dynamics for homogeneous systems and show how to extend existing data-based stabilization results to stabilize such error dynamics. The proposed method relies on the solution of a set of linear matrix inequalities that are shown to be feasible. Then, we solve the synchronization problem for heterogeneous systems by means of dynamic controllers. Different from existing results, we do not require model knowledge for the followers and the leader. The theoretical results are finally validated using a numerical simulation.

\end{abstract}

\section{INTRODUCTION}

An interesting area of research in control theory is the analysis and control of multiagent systems. This refers to the study of groups of dynamical agents that must interact with each other in order to achieve a given task. One of the most important applications of multiagent control is the solution of the synchronization problem, where agents that have limited sensing capabilities must synchronize their states to achieve a common value \cite{OlfatiMur2007}. Most of the solutions given to the synchronization problem are based on the exact knowledge of the dynamical models of the agents \cite{ZhangLeDa2011,WielandSepAll2011,ChowdhuryKha2021}.

In recent years, there has been a great interest in developing stabilizing controllers without the need of any knowledge of a system's model. One of the main drivers of this recent research has been Willems' fundamental lemma \cite{WillemsRapMarDe2005}, which provides a method to represent every input-output trajectory of a discrete-time (DT) linear system using only measured data. Continuous-time (CT) versions of this result have also been proposed \cite{LopezMuCDC2022,RapisardaCamWaa2023}. Willems' lemma has been recently leveraged for the data-based control of DT (see the survey \cite{MarkovskyDor2021}) and CT systems (see, e.g., \cite{RapisardavanCam2023,LopezMulAUT2024,EisingCor2023}).

Some results have been obtained for the data-driven control of multiagent systems. In \cite{Jiaoetal2021,Lietal2024} the synchronization problem is solved using only partial knowledge of the agents' dynamics. In \cite{ChangJiaLi2024}, consensus protocols are designed for systems with identical models using time-varying feedback matrices. Several iterative learning algorithms have been proposed to achieve synchronization as in, e.g., \cite{Modaresetal2016,ZhouLiGao2023}. In \cite{AllibhoyCor2021}, a multiagent predictive control scheme is used to minimize a global cost function.

In this paper, we study the data-based synchronization problem for continuous-time multiagent systems. In particular, we contribute with two novel solutions for this problem in different settings. First, we design distributed controllers that achieve synchronization in networked systems with homogeneous dynamics. The obtained controllers use time invariant feedback gains, and are static in the sense that no dynamic variables are required for control. Then, we solve the data-based synchronization problem for heterogeneous systems. This solution uses dynamic controllers that, different from the result in \cite{Jiaoetal2021,Lietal2024}, do not require any knowledge of the systems' models. To obtain these results, we provide a novel data-based representation of the synchronization error dynamics, leveraging the results in \cite{LopezMuCDC2022}. Different from the learning techniques in the literature \cite{Modaresetal2016,ZhouLiGao2023}, we provide a method to collect persistently exciting data that guarantees the feasibility of the proposed solutions. 

The paper is organized as follows. Relevant results from the literature are described in Section \ref{secprel}. Our solution to the synchronization problem in homogeneous systems is presented in Section \ref{sechomogen}, and our method for the heterogeneous case is shown in Section \ref{sechet}. We present a simulation example in Section \ref{secsimul}, and conclude the paper in Section \ref{secconc}.

\section{PRELIMINARIES}
\label{secprel}

\subsection{Notation}
\label{subsecnot}

The notation $P \succ 0$ denotes that $P$ is a symmetric positive definite matrix. An identity matrix of dimensions $n \times n$ is written as $I_n$. Given a sequence $\{ d_1,d_2,\ldots,d_M \}$, we represent a diagonal matrix with the values $d_i$ on the diagonal as $\diag_i\{ d_i \}$. Similarly, given a sequence of matrices $\{ D_1,\ldots, D_N \}$, $\blkdiag_i\{ D_i \}$ is a block diagonal matrix with diagonal blocks $D_i$. 

Consider the scalars $M \in \mathbb{N}$ and $T~\in~\mathbb{R}_+$. Let $\xi : [0, MT) \rightarrow \mathbb{R}^\sigma$ denote a continuous-time signal defined in the time interval $[0, MT)$. Using this signal, define the time-varying matrix
\begin{equation}
	\mathcal{H}_{T}(\xi(t)) := \left[ \begin{array}{cccc}
		\xi(t) & \xi(t+T) & \cdots & \xi(t+(M-1)T) \end{array} \right]
	\label{hankrow}
\end{equation}
for $0 \leq t < T$.

\subsection{Data-based representation and control of continuous-time systems}
\label{subsecrep}

Here, we summarize the developments in \cite{LopezMuCDC2022}. Consider a continuous-time (CT) linear system
\begin{subequations}
	\begin{align}
		\dot x &= A x + B u, \label{ctsysx}\\
		y & = C x + Du,
	\end{align}
	\label{ctsys}%
\end{subequations}
where $x \in \R^n$, $u \in \R^m$ and $y \in \R^p$ are the state, input and output vectors, respectively. Assume that input and output trajectories of \eqref{ctsys} are measured, and the matrices $\mH_T(u(t))$ and $\mH_T(y(t))$ are defined as in \eqref{hankrow}. In some cases, we assume that state and state derivative information is also available, with the corresponding matrices $\mH_T(x(t))$ and $\mH_T(\dot x(t))$. The assumption that $\dot x$ can be either measured or estimated has been made in several works in the literature \cite{BerberichWilHerAll2021,BisoffiPerTes2022}. This assumption can be relaxed by integrating the measured input-state trajectories instead (see, e.g., \cite[Appendix A]{PersisPosTes2022} and \cite{LopezMuCDC2023}). For simplicity of presentation, in this paper we let the trajectory $\dot x$ be available. 

For data-based system representation and control, it is useful to collect persistently excited (PE) data from the CT system. A class of PE inputs for \eqref{ctsys} is defined as follows. 

\begin{defn}[Piecewise constant PE input]
	\label{defctpe}
	Consider a time interval $T > 0$ such that
	\begin{equation}
		T \neq \frac{2 \pi \kappa}{| \mathcal{I}_m(\lambda_j - \lambda_k) |}, \qquad \forall \kappa \in \mathbb{Z},
		\label{assut}
	\end{equation}
	where $\mathcal{I}_m(\cdot)$ is the imaginary part of a complex number, and $\lambda_j$ and $\lambda_k$ are any two eigenvalues of the matrix $A$ in (\ref{ctsys}). A piecewise constant persistently exciting (PCPE) input of order $L$ is given by ${u(t + jT) = \mu_j}$, $0 \leq t < T$, $j=0,\ldots,M-1$, where the sequence $\{ \mu_j \}_{j=0}^{M-1}$, $\mu_j \in \mathbb{R}^m$, is such that
	\begin{equation*}
		\text{rank} \left( \left[ \begin{array}{cccc}
			\mu_0 & \mu_1 & \cdots & \mu_{M-L} \\
			\mu_1 & \mu_2 & \cdots & \mu_{M-L+1} \\
			\vdots & \vdots & \ddots & \vdots \\
			\mu_{L-1} & \mu_{L} & \cdots & \mu_{M-1}
		\end{array} \right] \right) = mL.
	\end{equation*}
\end{defn}

\vspace{0.1cm}
\begin{rem}
	Notice that, when a PCPE input is used, the matrix $\mathcal{H}_T(u(t))$ defined as in \eqref{hankrow} is constant for $0 \leq t < T$, and it is denoted simply as $\mathcal{H}_T(u)$.
\end{rem}

If a PCPE input is applied to a controllable system (\ref{ctsys}), then the corresponding input-state data satisfy an important property, as given by the following lemma.

\begin{lem}[\cite{LopezMuCDC2022}]
	\label{lempe}
	Consider system (\ref{ctsys}), let the pair $(A,B)$ be controllable, and let $u$ be a PCPE input of order $n+1$. Then,
	\vspace{-0.1cm}
	\begin{equation}
		\text{rank} \left( \left[ \begin{array}{c}
			\mathcal{H}_T(u) \\ \mathcal{H}_T(x(t))
		\end{array} \right] \right) = m + n
		\label{pecond}
	\end{equation}
	for all $0 \leq t < T$.
\end{lem}

In \cite{LopezMuCDC2022}, a data-based system representation for continuous-time systems was introduced, which corresponds to a continuous-time version of the well-known Willems' lemma \cite{WillemsRapMarDe2005}. The following theorem states this result.

\begin{thm}[\cite{LopezMuCDC2022}]
	\label{thmctwill}
	Consider a system (\ref{ctsys}) such that the pair $(A,B)$ is controllable. Let $u : [0,MT) \rightarrow \mathbb{R}^m$, $T > 0$, $M \in \mathbb{N}$, be a PCPE input of order $n+1$, and let $x : [0,MT) \rightarrow \mathbb{R}^n$ and $y : [0,MT) \rightarrow \mathbb{R}^p$, be the corresponding states and outputs of (\ref{ctsys}). Any signals $\bar u : [0,T) \rightarrow \mathbb{R}^m$, $\bar y : [0,T) \rightarrow \mathbb{R}^p$, where $\bar u$ is continuously differentiable, are an input-output trajectory of (\ref{ctsys}) corresponding to some initial condition $\bar x(0)$ if and only if they can be expressed as
	\begin{equation}
		\left[ \begin{array}{c}
			\mathcal{H}_T(u) \\ \mathcal{H}_T(y(t))
		\end{array} \right] \alpha(t) = \left[ \begin{array}{c}
			\bar u(t) \\ \bar y(t)
		\end{array} \right],
		\label{ctwillem}
	\end{equation}
	for $0 \leq t < T$, where $\alpha(t) \in \R^{M}$ satisfies
	\begin{equation}
		\left[ \begin{array}{c}
			\mathcal{H}_T(u) \\ \mathcal{H}_T(x(t))
		\end{array} \right] \dot \alpha(t) = \left[ \begin{array}{c}
			\dot{\bar{u}}(t) \\ 0
		\end{array} \right],
		\label{difeqa}
	\end{equation}
	\begin{equation}
		\left[ \begin{array}{c}
			\mathcal{H}_T(u) \\ \mathcal{H}_T(x(0))
		\end{array} \right] \alpha(0) = \left[ \begin{array}{c}
			\bar{u}(0) \\ \bar x(0)
		\end{array} \right].
		\label{inicoa}
	\end{equation}
\end{thm}

\vspace{0.1cm}
A straightforward consequence of Theorem \ref{thmctwill} is that the state trajectory $\bar x$ that corresponds to $(\bar u, \bar y)$ (the state realization is given by the data $x$ in \eqref{difeqa}) can be written as
\begin{equation}
	\bar x(t) = \mH_T(x(t)) \alpha(t).
	\label{barx}
\end{equation}
\vspace{-0.1cm}
Taking the time derivative of \eqref{barx}, and noticing that $\mH_T(x(t)) \dot \alpha(t) = 0$ as per \eqref{difeqa}, we also obtain
\begin{equation}
	\dot{\bar{x}}(t) = \mH_T(\dot x(t)) \alpha(t).
	\label{barxdot}
\end{equation}

An important application of Theorem \ref{thmctwill} regards solving the data-based simulation problem, which consists of using measured data to determine the output trajectory $\bar y$ of \eqref{ctsys} for a given input trajectory $\bar u$ and an initial condition $\bar x_0$. In this paper, we make use of a particular case of the simulation problem, where $y=x$. Suppose that the desired trajectory to simulate is $\bar u \equiv 0$. Then, it is clear that $\dot \alpha \equiv 0$ satisfies \eqref{difeqa}. Then, the desired state trajectory is given simply by
\begin{equation}
	\bar x(t) = \mH_T(x(t)) \alpha(0), \quad 0 \leq t < T.
	\label{simprob}
\end{equation}

We conclude this subsection by commenting on control design. Different data-based controllers for CT systems have been proposed \cite{RapisardavanCam2023,LopezMulAUT2024}. A simple stabilizing controller was described in \cite[Remark~2]{DePersisTes2020} as follows. The matrix $K = -\mH_T(u(0)) \Lambda \left( \mH_T(x(0)) \Lambda \right)^{-1}$, where $\Lambda \in \R^{M \times n}$ satisfies
\begin{gather}
	\mathcal{H}_T(x(0)) \Lambda \succ 0, \label{deper1}\\
	\mathcal{H}_T(\dot x(0)) \Lambda + \Lambda^\top \mathcal{H}_T(\dot x(0))^\top \prec 0,
	\label{deper2}
\end{gather}
is such that $A-BK$ is Hurwitz, i.e., system \eqref{ctsys} is stabilized.

\subsection{Graph theory definitions}
\label{subsecgraph}

Consider a directed graph $\mathcal{G} = (\mathcal{V},\mathcal{E})$, where $\mathcal{V}= \{ v_i \}_{i=1}^N$ is a set of nodes and $\mathcal{E} \subseteq \mathcal{V} \times \mathcal{V}$ is a set of edges. In this paper, the nodes represent a set of $N$ dynamical agents and the edges correspond to their communication links. Define the graph weights $a_{ij}$, $i,j = 1, \ldots, N$, with $a_{ij} > 0$ if $(v_j,v_i) \in \mathcal{E}$ and $a_{ij}=0$ otherwise. The graph is assumed to have no self-loops, i.e., $a_{ii} = 0$ for all agents $i$. The adjacency matrix $\mathcal{A} \in \mathbb{R}^{N \times N}$ of the graph $\mathcal{G}$ is $\mathcal{A} = [ a_{ij} ]$. The weighted in-degree of each node $i$ is given by $d_i = \sum_{j=1}^N a_{ij}$, and the in-degree matrix of $\mathcal{G}$ is $D=\diag_i\{ d_i \}$. The Laplacian matrix of the graph is finally given by $\mathcal{L} = D-\mathcal{A}$.

Additional to the $N$ agents described above, consider an additional node, $v_0$, regarded as the \emph{leader} node. The pinning gain $g_i$ is such that $g_i>0$ if agent $i$ can access the leader information; otherwise $g_i = 0$. The pinning matrix is defined as $G= \diag_i\{ g_i \}$. For convenience, we also defined the pinning vector $g \in \mathbb{R}^N$ as $g = [g_1 \cdots g_N]^\top$. 

The graph $\mathcal{G}$ is said to have a spanning tree if there exists at least one path of directed edges connecting a node, called the \emph{root}, with every other node in $\mathcal{G}$. The following assumption will be made throughout the paper.

\begin{assum}
	\label{asssp}
	The graph $\mathcal{G}$ has a spanning tree, and the leader is connected to a root node.
\end{assum}

\subsection{The synchronization problem for homogeneous agents}

Consider a set of $N$ dynamical systems with homogeneous (i.e., identical) linear dynamics as
\begin{equation}
	\dot x_i = A x_i + B u_i, \quad i=1,\ldots,N,
	\label{agenti}
\end{equation}
where $x_i \in \mathbb{R}^n$ and $u_i \in \mathbb{R}^m$ are the state and input vectors of agent $i$, respectively. The leader dynamics are given by
\begin{equation}
	\dot x_0 = A x_0,
	\label{leader}
\end{equation}
where $x_0 \in \mathbb{R}^n$ is the leader state. The $N$ agents can only access their own state information and that of their close neighbors as described by a communication graph $\mathcal{G}$.

The synchronization problem is formulated as follows.

\begin{prob}
	\label{pr1}
	Consider the multiagent system \eqref{agenti}-\eqref{leader}. Design distributed control inputs $u_i$ for all agents such that state synchronization with the leader is achieved, that is, $x_i(t) - x_0(t) \rightarrow 0$ as $t \rightarrow \infty$ for $i=1,\ldots,N$.
\end{prob}

\subsection{The synchronization problem for heterogeneous agents}

Consider a heterogeneous multiagent system as
\begin{subequations}
	\begin{align}
		\dot x_i &= A_i x_i + B_i u_i, \\
		y_i & = C_i x_i	\quad i=1,\ldots,N,
	\end{align}
\label{heteroi}%
\end{subequations}
where the state and input dimensions may differ between the systems, $x_i \in \mathbb{R}^{n_i}$, $u_i \in \mathbb{R}^{m_i}$. The output dimensions are the same for all agents, $y_i \in \mathbb{R}^p$. The leader has the dynamics
\begin{subequations}
	\begin{align}
		\dot x_0 & = A_0 x_0, \label{hetleadx}\\
		y_0 & = C_0 x_0, \label{hetleady}
	\end{align}
	\label{heterol}%
\end{subequations}
with $x_0 \in \mathbb{R}^{n_0}$, $y_0 \in \mathbb{R}^p$.

\begin{rem}
	\label{remxy}
	Although the systems \eqref{heteroi}-\eqref{heterol} have outputs $y_i$, it is a common assumption in the synchronization literature that the complete state $x_i$ is measurable (see, e.g., \cite{ZhangLeDa2011,Jiaoetal2021}). Rather than measurements, the outputs $y_i$ represent the signals to be synchronized. If only output measurements are available as in, e.g., \cite{WielandSepAll2011}, an observer can be included in the control loop. We do not consider this setting in this paper. 
\end{rem}

Our interest now is to solve the following problem.

\begin{prob}
	\label{pr2}
	Consider the multiagent system \eqref{heteroi}-\eqref{heterol}. Design distributed control inputs $u_i$ for all agents such that the output synchronization error satisfies $y_i(t) - y_0(t) \rightarrow 0$ as $t \rightarrow \infty$ for $i=1,\ldots,N$.
\end{prob}

We now present our data-based solutions for Problems~\ref{pr1} and \ref{pr2}. Our reason for solving these problems separately is that, different from our solution to Problem~2, we solve Problem~\ref{pr1} without the use of dynamic variables for control. This latter case is described in the following section.

\section{DATA-BASED SYNCHRONIZATION OF HOMOGENEOUS AGENTS}
\label{sechomogen}

Our approach to solve Problem \ref{pr1} is to formulate an LMI-based procedure similar to \eqref{deper1}-\eqref{deper2}, but placing additional constraints to account for the multiagent nature of the problem. First, we show a useful fact about a model-based solution of Problem \ref{pr1} that will be used in later subsections.

\subsection{A result on model-based synchronization control}
\label{secres}

Consider the homogeneous multiagent system \eqref{agenti}-\eqref{leader} and define the local synchronization error of agent $i$ as
\begin{equation}
	\delta_i = \sum_{j=1}^N a_{ij} (x_i - x_j) + g_i (x_i - x_0).
	\label{deltae}
\end{equation}
Using the dynamics \eqref{agenti}-\eqref{leader}, the synchronization error dynamics are given by
\begin{equation}
	\dot \delta_i = A \delta_i + \sum_{j=1}^N a_{ij} B (u_i - u_j) + g_i B u_i
	\label{deltaedot}
\end{equation}
and the global error dynamics are
\begin{equation}
	\dot \delta = (I \otimes A) \delta + ((\mL + G) \otimes B) u
	\label{globaldelta}
\end{equation}
where $\otimes$ represents the Kronecker product, and
\begin{equation}
	\delta = [\delta_1^\top \quad \delta_2^\top \; \cdots \; \delta_N^\top]^\top \in \R^{nN},
	\label{deltavec}
\end{equation}
\begin{equation}
	u = [u_1^\top \quad u_2^\top \; \cdots \; u_N^\top]^\top \in \R^{mN}.
	\label{uvec}
\end{equation}

It is well known \cite{ZhangLeDa2011} that the distributed control inputs
\begin{equation}
	u_i = -K \delta_i, 
	\label{inputk}
\end{equation}
$i=1,\ldots,N$, solve Problem \ref{pr1} if and only if the gain $K \in \R^{m \times n}$ is such that the following matrix is Hurwitz
\begin{equation}
	A_c := (I \otimes A) - ((\mL + G) \otimes BK).
	\label{ac}
\end{equation}
It was shown in \cite{ZhangLeDa2011} that $K$ can be designed by selecting 
\begin{equation}
	K= c R^{-1} B^\top \bar P,
	\label{ck}
\end{equation}
where $c$ is a large enough scalar, $\bar P \in \R^{n \times n}$ is the solution of the algebraic Riccati equation (ARE)
\begin{equation}
	Q + \bar PA + A^\top \bar P - \bar PBR^{-1}B^\top \bar P =0, 
	\label{riccati}
\end{equation}
and $Q \in \mathbb{R}^{n \times n}$ and $R \in \mathbb{R}^{m \times m}$ are user-defined positive definite matrices. 

In the following lemma we show that, if $c$ is sufficiently large, then we can always find a \emph{block diagonal} matrix $P =\blkdiag_i\{ P_i \}$, for some $P_i \succ 0$, $i=1,\ldots,N$, such that 
\begin{equation}
	P A_c + A_c^\top P \prec 0.
	\label{lyapac}
\end{equation}

\begin{lem}
	\label{lembkdiag}
	Consider the multiagent system \eqref{agenti}-\eqref{leader}, let the pair $(A,B)$ be controllable, and let Assumption~\ref{asssp} hold. There exists a matrix $K \in \mathbb{R}^{m \times n}$ and a block diagonal matrix $P = \blkdiag_i\{ P_i \}$, $P_i \in \mathbb{R}^{n \times n}$, $i=1,\ldots,N$, such that $P \succ 0$ and \eqref{lyapac} holds with $A_c$ as in \eqref{ac}.
\end{lem}
\begin{proof}
	It was shown in \cite[Theorem 1]{Zhangetal2015} that, under Assumption~\ref{asssp}, there exists a diagonal matrix $S = \diag_i\{ s_i \}$, $s_i > 0$, $i=1,\ldots,N$, such that $(\mathcal{L} + G) S + S (\mathcal{L} + G) = \bar Q \succ 0$. Observe that, since $\bar Q \succ 0$, there exists a small enough constant $\epsilon > 0$ such that $\bar Q - \epsilon I \succ 0$.
	
	Now, let $K$ be as in (\ref{ck})-(\ref{riccati}) and define $P=\blkdiag_i\{ P_i \}:=\blkdiag_i\{ s_i \bar P \}$. Hence,
	\begin{equation*}
		P A_c = \blkdiag_i\{ s_i \bar P \} [(I \otimes A) - c((\mathcal{L} + G) \otimes BR^{-1}B^\top \bar P)].
	\end{equation*}
	Using $(\mathcal{L} + G) S + S (\mathcal{L} + G) = \bar Q$, we have
	\begin{multline*}
		P A_c + A_c^\top P = \blkdiag_i\{ s_i \left(\bar P A + A^\top \bar P \right) \} \\
		- c \left( \bar Q \otimes \bar PBR^{-1}B^\top \bar P \right).
	\end{multline*}
	Adding and subtracting $\epsilon c (I \otimes \bar P BR^{-1} B^\top \bar P)$, we get
	\begin{multline*}
		P A_c + A_c^\top P = - c \left( (\bar Q - \epsilon I) \otimes \bar PBR^{-1}B^\top \bar P \right) \\
		+ \blkdiag_i\{ s_i \left(\bar P A + A^\top \bar P - (\epsilon c/s_i) \bar PBR^{-1}B^\top \bar P \right) \}.
	\end{multline*}
	We have said that $\bar Q - \epsilon I \succ 0$, which implies that the first term on the right-hand side of this equation is negative semidefinite. We complete the proof by showing that $\bar P A + A^\top \bar P - (\epsilon c/s_i) \bar PBR^{-1}B^\top \bar P \prec 0$ for $i=1,\ldots,N$. Using (\ref{riccati}), notice that 
	\begin{multline*}
		\bar P A + A^\top \bar P - (\epsilon c/s_i) \bar PBR^{-1}B^\top \bar P \\ = -Q-((\epsilon c/s_i) -1)\bar P B R^{-1}B^\top \bar P,
	\end{multline*}
	which is negative definite for a large enough $c>0$.
\end{proof}

Lemma \ref{lembkdiag} will be used below to solve the data-based synchronization problem. In the following subsection, we obtain a data-based characterization of the synchronization error dynamics to be stabilized.

\subsection{Data-based representation of the synchronization error dynamics}
\label{subsecrephom}

In this subsection, we use the framework described in \cite{LopezMuCDC2022} (and summarized in Section \ref{subsecrep} above) to represent the CT synchronization error trajectories in a data-based fashion.

Since the $N$ systems \eqref{agenti} have identical models, it is enough to collect PE data from one of them to generate the global system trajectories. Thus, take one of the agents \eqref{agenti}, apply a PCPE input of order $n+1$, and collect the input-state trajectories $(u,x)$. According to Theorem~\ref{thmctwill}, there exists a vector $\alpha_i(t)$ such that every input-state trajectory $(\bar u_i,\bar x_i)$ of agent $i$ in (\ref{agenti}) can be expressed as
\begin{equation}
	\begin{bmatrix}
		\mathcal{H}_T (u) \\ \mathcal{H}_T (x(t))
	\end{bmatrix} \alpha_i(t) = \begin{bmatrix}
	\bar u_i(t) \\ \bar x_i(t)
\end{bmatrix}
\label{dbagei}
\end{equation}
for $0 \leq t < T$. Any leader trajectory $\bar x_0$ can also be obtained as $\mH_T(x(t)) \alpha_0(t) = \bar x_0(t)$,
where $\alpha_0$ also satisfies 
\begin{equation}
	\mH_T(u) \alpha_0(t) = 0, \quad 0\leq t < T,
	\label{ualpha0}
\end{equation}
to account for the unforced dynamics in \eqref{leader}.

Replacing $x_i$ by $\bar x_i$ in (\ref{deltae}) and substituting $\bar x_i(t) = \mH_T(x(t)) \alpha_i(t)$ from \eqref{dbagei}, we obtain
\begin{multline}
	\delta_i(t) =  \mathcal{H}_T (x(t)) \Bigl[ \sum_{j=1}^N a_{ij} \left( \alpha_i(t) - \alpha_j(t) \right) \\
	+ g_i \left( \alpha_i(t) - \alpha_0(t) \right) \Bigr].
	\label{deltaidata}
\end{multline}
We can write \eqref{deltaidata} in global form by defining $\delta$ as in \eqref{deltavec} and $\alpha = [\alpha_0^\top \,\, \alpha_1^\top \cdots \alpha_N^\top]^\top \in \R^{M(N+1)}$. This yields
\begin{equation}
	 \delta(t) = \bigl( [-g \quad \mathcal{L}+G] \otimes \mathcal{H}_T (x(t)) \bigr) \alpha,
	 \label{dbdelta}
\end{equation}
where $\mL$, $g$ and $G$ were defined in Section \ref{subsecgraph}. 

For the dynamics of $\delta$, from \eqref{dbdelta} and the fact that $\mH_T(x(t))\dot{\alpha}_i(t)=0$ for all $i=0,\ldots,N$ from \eqref{difeqa}, we obtain
\begin{equation}
	\dot \delta(t) = \bigl( [-g \quad \mathcal{L}+G] \otimes \mathcal{H}_T(\dot x(t)) \bigr) \alpha.
	\label{dbdeltadot}
\end{equation}

Finally, since $\bar u_i = \mH_T(u) \alpha_i$ as in \eqref{dbagei} for any input $\bar u_i$, the global input $u$ as in \eqref{uvec} is
\begin{equation}
	u = ([0_{N \times 1} \quad I_N] \otimes \mH_T(u)) \alpha.
	\label{dbinput}
\end{equation}

\subsection{Stabilization of the synchronization error}
\label{subsecstab}

Taking advantage of the representations in \eqref{dbdelta} and \eqref{dbdeltadot}, we can use any data-based controller in the literature to stabilize the synchronization error. In the following, we use the procedure in \eqref{deper1}-\eqref{deper2}. We begin by showing the following lemma. For convenience, define the matrices $\mL_g = [-g \quad \mathcal{L}+G]$ and $\mathcal{I}_g = [0_{N \times 1} \quad I_N]$.

\begin{lem}
	\label{lemmat}
	Consider the homogeneous multiagent system \eqref{agenti}-\eqref{leader} where the pair $(A,B)$ is controllable, and let Assumption \ref{asssp} hold. Apply a PCPE input of order $n+1$ to any one of the agents and collect the input-state data $(u,x)$. The following matrix has full row rank
	\begin{equation}
		\begin{bmatrix}
			I_{N+1} \otimes \mH_T(u) \\ \mL_g \otimes \mH_T(x(0))
			\label{HxHu}
		\end{bmatrix}.
	\end{equation}
\end{lem}
\vspace{0.1cm}
\begin{proof}
	By Assumption \ref{asssp}, $\mL + G$ is nonsingular \cite{Zhangetal2015}, and hence $\mL_g$ has full row rank. By persistence of excitation, \eqref{pecond} holds for all $0 \leq t < T$ \cite{LopezMuCDC2022}. It is well-known that, by the properties of the Kronecker product, $\rank(\mL_g \otimes \mH_T(x(0))) = \rank(\mL_g) \rank(\mH_T(x(0)))$ \cite{Tian2005}, and therefore both $\mL_g \otimes \mH_T(x(0))$ and $I_{N+1} \otimes \mH_T(u)$ have full row rank. The proof is completed by noticing that, by definition of the Kronecker product and by the full row rank nature of $I_{N+1}$ and $\mL_g$, a vector $v \neq 0$ in the left kernel of  \eqref{HxHu} implies the existence of a non-trivial vector on the left kernel of the matrix in \eqref{pecond}, contradicting Lemma \ref{lempe}.
\end{proof}

A direct application of \eqref{deper1}-\eqref{deper2} to the data-based representations \eqref{dbdelta} and \eqref{dbdeltadot} provides a stabilizing global controller, as we show in the following lemma. However, this is not a proper solution to Problem \ref{pr1} because, in general, distributed controllers as in \eqref{inputk} are not obtained.

\begin{lem}
	\label{lemdp}
	Let the conditions in Lemma \ref{lemmat} hold. Determine a matrix $\Lambda \in \mathbb{R}^{M(N+1) \times nN}$ such that
	\begin{subequations}
		\begin{gather}
			(\mathcal{L}_g  \otimes \mathcal{H}_T(x(0))) \Lambda \succ 0, \label{homdeper1}\\
			(\mathcal{L}_g  \otimes \mathcal{H}_T(\dot x(0))) \Lambda + \Lambda^\top (\mathcal{L}_g  \otimes \mathcal{H}_T(\dot x(0)))^\top \prec 0, \label{homdeper2} \\
			\left[ \mathcal{H}_T(u) \quad 0_{m \times MN} \right] \Lambda = 0. \label{homdeper3}
		\end{gather}
		\label{homdeper}%
	\end{subequations}
	The problem \eqref{homdeper} is feasible. Moreover, the input $u=-K\delta$ with $K=-(\mathcal{I}_g \otimes \mH_T(u)) \Lambda ((\mL_g \otimes \mH_T(x(0))) \Lambda)^{-1}$ renders the global error dynamics \eqref{globaldelta} asymptotically stable.
\end{lem}
\begin{proof}
If $u$ is selected as $u= -K \delta$, then from \eqref{globaldelta},
\begin{equation}
	\dot \delta = \left[ (I \otimes A) -((\mL +G) \otimes B) K \right] \delta.
	\label{prooflemdp1}
\end{equation}
Denote $A_\mL := \left[ (I \otimes A) -((\mL +G) \otimes B) K \right]$. Since \eqref{prooflemdp1} holds for any $\delta \in \R^{nN}$, consider a set of $nN$ arbitrary values of $\delta$ denoted as $\delta^k$, $k=1,\ldots,nN$, and write
\begin{equation*}
	\left[ \dot \delta^1 \quad \dot \delta^2 \; \cdots \; \dot \delta^{nN} \right] = A_\mL \left[ \delta^1 \quad \delta^2 \; \cdots \; \delta^{nN} \right].
\end{equation*}
Notice that each $\delta^k \in \R^{nN}$ can be expressed in data-based fashion as in \eqref{dbdelta}, i.e., $\delta^k= (\mL_g \otimes \mH_T(x(0)))\alpha^k$ with $\alpha^k \in \R^{M(N+1)}$, and similarly for $\dot \delta^k$ using \eqref{dbdeltadot}. Defining $\Lambda := \left[ \alpha^1 \quad \alpha^2 \; \cdots \; \alpha^{nN} \right] \in \R^{M(N+1) \times nN}$, we can write
\begin{equation}
	(\mL_g \otimes \mH_T(\dot x)) \Lambda = A_\mL (\mL_g \otimes \mH_T(x)) \Lambda.
	\label{prooflemdp2}
\end{equation}
Thus, giving values to the matrix $\Lambda$ can be understood as determining vectors $\alpha^k_i$ that generate closed-loop system trajectories for each agent $i$. Note that \eqref{ualpha0} must hold to have a proper data-based representation of the multiagent system. Hence, we restrict the first $N$ rows of $\Lambda$ (i.e., those corresponding to $\alpha_0$) to satisfy \eqref{ualpha0}, as given by \eqref{homdeper3}.
   
Suppose that, as required by the LMI \eqref{homdeper1}, there exists a value of $\Lambda$ such that $(\mL_g \otimes \mH_T(x)) \Lambda =P$ for some $P \succ 0$ (this is always possible, as shown below). Then, from \eqref{prooflemdp2} we have $A_\mL = (\mL_g \otimes \mH_T(\dot x)) \Lambda P^{-1}$. From this expression, we conclude that the LMI \eqref{homdeper2} is equivalent to $A_\mL P + P A_\mL \prec 0$. This implies that $K$ stabilizes the closed-loop system \eqref{prooflemdp1}.

Since we apply the input $u= -K \delta$ to the system \eqref{globaldelta}, then from \eqref{dbdelta} and \eqref{dbinput}, we write $(\mathcal{I}_g \otimes \mH_T(u)) \Lambda = -K (\mL_g \otimes \mH_T(x)) \Lambda$. This expression leads us to the matrix $K$ in the lemma statement.

Finally, we show that we can always select $\Lambda$ such that 
\begin{equation*}
	\begin{bmatrix}
		[ \mH_T(u) \quad 0_{m \times MN}] \\ \mathcal{I}_g \otimes \mH_T(u) \\ \mL_g \otimes \mH_T(x(0))
	\end{bmatrix} \Lambda = \begin{bmatrix}
	0 \\ -K P \\ P
\end{bmatrix}.
\end{equation*}
The proof is given by the fact that the left-most matrix in this expression is equal to the matrix \eqref{HxHu} which, by Lemma~\ref{lemmat}, has full row rank.
\end{proof}

Lemma \ref{lemdp} provides a global input of the form $u = -K \delta$ which does not necessarily correspond to the distributed inputs required in Problem \ref{pr1} (see, e.g., \eqref{inputk}). In the following theorem, we show that this issue can be solved with additional constraints in the LMIs \eqref{homdeper}.

\begin{thm}
	\label{thmhom}
	Let the conditions in Lemma \ref{lemmat} hold. Consider the problem of determining matrices $\Lambda \in \mathbb{R}^{M(N+1) \times nN}$, $P_i \in \mathbb{R}^{n \times n}$ and $F_i \in \mathbb{R}^{m \times n}$, $i=1\ldots,N$, such that
	\begin{subequations}
		\begin{gather}
			(\mathcal{L}_g  \otimes \mathcal{H}_T(x(0))) \Lambda = \blkdiag_i\{P_i\}, \label{homlgx}\\
			(\mathcal{L}_g  \otimes \mathcal{H}_T(\dot x(0))) \Lambda + \Lambda^\top (\mathcal{L}_g  \otimes \mathcal{H}_T(\dot x(0)))^\top \prec 0, \label{homlgdx} \\
			\left[ \mathcal{H}_T(u) \quad 0_{m \times MN} \right] \Lambda = 0, \label{hom0} \\
			P_i \succ 0, \quad i=1,\ldots,N, \label{homp}\\
			(\mathcal{I}_g  \otimes \mathcal{H}_T(u)) \Lambda = \blkdiag_i\{ F_i \}. \label{homf}
		\end{gather}
		\label{homsol}%
	\end{subequations}
The problem \eqref{homsol} is feasible. Moreover, the distributed inputs $u_i = - K_i \delta_i$ with $K_i = F_i P_i^{-1}$, $i=1,\ldots,N$, solve Problem~\ref{pr1}.
\end{thm}
\begin{proof}
	From Lemma \ref{lemdp} it follows that \eqref{homlgx}-\eqref{homp} provide a stabilizing global input of the form $u = -K\delta$, where $K = - (\mathcal{I}_g \otimes \mH_T(u)) \Lambda ((\mL_g \otimes \mH_T(x(0))) \Lambda)^{-1}$. The conditions \eqref{homlgx} and \eqref{homf} constrain the matrices to be block diagonal, such that $K = - \blkdiag_i\{ F_i \} \blkdiag_i\{P_i^{-1}\} = \blkdiag_i\{ K_i \}$ with $K_i$ as in the theorem statement.
	
	The proof of feasibility is given by Lemma \ref{lembkdiag}, which shows the existence of a block diagonal matrix $K$ for which there exists a block diagonal matrix $P$ such that \eqref{lyapac} holds.
\end{proof}

In the following section, we address the synchronization problem for heterogeneous agents as described in Problem~\ref{pr2}.

\section{DATA-BASED SYNCHRONIZATION OF HETEROGENEOUS SYSTEMS}
\label{sechet}

In \cite{WielandSepAll2011}, sufficient and necessary conditions to solve the synchronization problem of heterogeneous agents without a leader are provided. There, the following standard assumption is made\footnote{The controllers presented in this section also solve Problem \ref{pr2} if $A_0$ has unstable eigenvalues under additional assumptions on the connections of the communication graph $\mathcal{G}$. See the discussion in \cite[Section~3.4]{WielandSepAll2011}.}.

\begin{assum}
	\label{assumeig}
	The matrix $A_0$ in \eqref{heterol} has all its eigenvalues on the imaginary axis.
\end{assum}

Using similar arguments as in \cite{WielandSepAll2011} (see also \cite{Modaresetal2016}), it can be easily shown that the dynamic controllers
\begin{gather}
	\dot \zeta_i = A_0 \zeta_i - \sum_{j=1}^N a_{ij} (\zeta_i - \zeta_j)-g_i(\zeta_i - x_0), \label{dynamicu1} \\
	u_i = -K_i (x_i - \Pi_i \zeta_i) + \Gamma_i \zeta_i,
	\label{dynamicu2}
\end{gather}
$i = 1,\ldots,N$, with $K_i$ such that $A_i-B_iK_i$ is Hurwitz, solve the leader-follower synchronization setting of Problem~\ref{pr2} if and only if there exist matrices $\Pi_i \in \mathbb{R}^{n_i \times n_0}$ and $\Gamma_i \in \mathbb{R}^{m_i \times n_0}$, for all agents $i$, such that
\begin{gather}
		A_i \Pi_i + B_i \Gamma_i = \Pi_i A_0, \label{hetcon1}\\
		C_i \Pi_i = C_0.
	\label{hetcon2}
\end{gather} 

In this section, we propose to use a dynamic controller as in \eqref{dynamicu1}-\eqref{dynamicu2} designed entirely from measured data. A discrete-time version of this problem was studied in \cite{Jiaoetal2021} and \cite{Lietal2024}, where the equations \eqref{hetcon1}-\eqref{hetcon2} are solved by leveraging data matrices. However, those solutions require partial knowledge of the systems model. In particular, the matrices $C_i$ must be known for all agents $i$, and complete knowledge of the leader dynamics \eqref{heterol} is needed. In this paper, we provide a solution that does not require any knowledge of the systems models \eqref{heteroi}-\eqref{heterol}.

\subsection{Data-based representation of heterogeneous dynamics}

As in Section \ref{sechomogen}, it will be useful to obtain data-based representations of the dynamical systems of interest. However, in this section we are not concerned with representing the synchronization errors, but only the system trajectories from \eqref{heteroi} and \eqref{heterol}.

First, we consider $N$ dynamical systems with (potentially) different linear models as in \eqref{heteroi}. Since the models are different, data must be collected from each agent individually. Thus, assuming that all pairs $(A_i,B_i)$ are controllable, apply a PCPE input of order $n_i+1$ to each agent $i$, and collect the data $(u_i,x_i,y_i)$ (see Remark~\ref{remxy} regarding the availability of both state and output data). Now, using \eqref{ctwillem} and \eqref{barx}, we represent every input-state-output trajectory $(\bar u_i,\bar x_i,\bar y_i)$ of agent $i$ for $0 \leq t < T$ as
\begin{equation}
	\begin{bmatrix}
		\mathcal{H}_T (u_i) \\ \mathcal{H}_T (x_i(t)) \\ \mathcal{H}_T (y_i(t))
	\end{bmatrix} \alpha_i(t) = \begin{bmatrix}
		\bar u_i(t) \\ \bar x_i(t) \\ \bar y_i(t)
	\end{bmatrix}.
\end{equation} 

For the leader \eqref{heterol}, we must assume that we measure a state trajectory $x_0$ such that $\rank(\mH_T(x_0(t)))~=~n_0$ for all $0 \leq t < T$, without the need of a PE input\footnote{Compare this assumption with the results in \cite{MarkovskyDor2023}, where the conditions of the discrete-time Willems' lemma for data-based representation of system trajectories were relaxed for the case of uncontrollable systems.}. Using this state information and the corresponding output data from \eqref{hetleady}, every leader state-output trajectory is expressed as
\begin{equation*}
	\begin{bmatrix}
		\mathcal{H}_T (x_0(t)) \\ \mathcal{H}_T (y_0(t))
	\end{bmatrix} \alpha_0(t) = \begin{bmatrix}
		\bar x_0(t) \\ \bar y_0(t)
	\end{bmatrix}.
\end{equation*}

In this section, it is useful to notice that, from the system dynamics \eqref{heteroi}-\eqref{heterol}, the equations
\begin{subequations}
	\begin{gather}
		\mH_T(\dot x_i(t)) = A_i \mH_T(x_i(t)) + B_i \mH_T(u_i), \label{Hxdyni}\\
		\mH_T(y_i(t)) = C_i \mH_T(x_i(t)), \label{Hydyni}
	\end{gather}
\end{subequations}
$i=1,\ldots,N$, as well as
\begin{subequations}
	\begin{gather}
		\mH_T(\dot x_0(t)) = A_0 \mH_T(x_0(t)), \label{Hxdyn0}\\
		\mH_T(y_0(t)) = C_0 \mH_T(x_0(t)), \label{Hydyn0}
	\end{gather}
\end{subequations}
hold for $0 \leq t < T$.

\subsection{Synchronization of heterogeneous systems}

The solution to Problem \ref{pr2} that we develop in this paper consists of a purely data-based version of the method in \eqref{dynamicu1}-\eqref{dynamicu2}. We begin by using the system representation described in the previous subsection so solve the equations \eqref{hetcon1}-\eqref{hetcon2} in a data-based fashion.

Consider any of the agents $i$. By persistence of excitation (see \eqref{pecond}), we can always determine a matrix $S_i \in \R^{M \times n_0}$ such that 
\begin{equation}
	\begin{bmatrix}
		\mH_T(u_i) \\ \mH_T(x_i(0))
	\end{bmatrix} S_i = \begin{bmatrix}
	\Gamma_i \\ \Pi_i
\end{bmatrix}
\label{gampi}
\end{equation}
for any $\Gamma_i$, $\Pi_i$. Substituting in \eqref{hetcon1}, we get 
\begin{equation*}
	A_i \mH_T(x_i(0)) S_i + B_i \mH_T(u_i) S_i = \mH_T(x_i(0)) S_i A_0.
\end{equation*}
Multiplying this expression on the right by $\mH_T(x_0(0))$ and using \eqref{Hxdyni} and \eqref{Hxdyn0}, we obtain
\begin{equation}
	\mH_T(\dot x_i(0)) S_i \mH_T(x_0(0)) = \mH_T(x_i(0)) S_i \mH_T(\dot x_0(0)).
	\label{dbhetcon1}
\end{equation}
Following a similar procedure, substitute \eqref{gampi} in \eqref{hetcon2} and post-multiply by $\mH_T(x_0(0))$ to obtain $C_i \mH_T(x_i(0))S_i \mH_T(x_0(0))=C_0 \mH_T(x_0(0))$. From \eqref{Hydyni} and \eqref{Hydyn0}, we get
\begin{equation}
	\mH_T(y_i(0)) S_i \mH_T(x_0(0)) = \mH_T(y_0(0)).
	\label{dbhetcon2}
\end{equation}
 Now, \eqref{dbhetcon1} and \eqref{dbhetcon2} are data-based equations that can be solved for the unknown $S_i$. After computing $S_i$, the desired matrices $\Pi_i$ and $\Gamma_i$ are obtained from \eqref{gampi}.

The stabilizing matrix $K_i$ in \eqref{dynamicu2} can be computed using only data by means of known methods as, for example, \eqref{deper1}-\eqref{deper2}. Therefore, only the first term of \eqref{dynamicu1} remains containing model information. To solve this issue, recall that we have a method to generate a trajectory with dynamics of the form \eqref{hetleadx} by means of the expression \eqref{simprob} that uses a constant $\alpha$, as described in Section \ref{subsecrep}. Thus, given the initial condition $\zeta_i(0)$, compute a constant vector $\bar \alpha_i$ such that $\mH_T(x_0(0)) \bar \alpha_i= \zeta_i(0)$. Then, since $A_0 \zeta_i(0) = A_0 \mH_T(x_0(0)) \bar \alpha_i$, replace \eqref{dynamicu1} with
\begin{multline}
	\dot \zeta_i(t) = \mH_T(\dot x_0(t)) \bar \alpha_i - \sum_{j=1}^N a_{ij} (\zeta_i (t) - \zeta_j(t)) \\
	-g_i(\zeta_i(t) - x_0(t)).
	\label{dbdynamicu}
\end{multline}

\begin{rem}
	The operation $\mH_T(\dot x_0(t)) \bar \alpha_i$ can only generate trajectories in the time interval $0 \leq t < T$. However, this is not a restriction in the applicability of the proposed method. When $t = kT$ for any $k \in \mathbb{N}$, recompute $\bar \alpha_i$ such that $\mH_T(x_0(0)) \bar \alpha_i= \zeta_i(kT)$, and restart the index $t$ of the data matrix $\mH_T(\dot x_0(t))$ in \eqref{dbdynamicu}.
\end{rem}

The proposed data-based method to solve Problem \ref{pr2} is formalized in the following theorem.

\begin{thm}
	\label{thmhet}
	Consider the heterogeneous multiagent system \eqref{heteroi}-\eqref{heterol} where every pair $(A_i,B_i)$, $i=1,\ldots,N$, is controllable, and let Assumptions \ref{asssp} and \ref{assumeig} hold. Apply a PCPE input of order $n_i+1$ to each agent $i$ and collect the input-state-output data $(u_i,x_i,y_i)$. Moreover, collect a leader state-output trajectory $(x_0,y_0)$ such that $\rank(\mH_T(x_0(t))) = n_0$ for all $0 \leq t < T$. Let $K_i$ be matrices such that $A_i-B_i K_i$ are Hurwitz for all agents $i$. There exist distributed dynamic controllers \eqref{dynamicu2} with $\zeta_i$ satisfying \eqref{dbdynamicu} that solve Problem \ref{pr2} if and only if there is a matrix $S_i$ that satisfies the matrix equations \eqref{dbhetcon1}-\eqref{dbhetcon2}.
\end{thm}
\begin{proof}
	By construction and the full-rank property of $\mH_T(u_i),\mH_T(x_i(0))$, $i=1,\ldots,N$, and of $\mH_T(x_0(0))$, the data-based equations \eqref{dbhetcon1}-\eqref{dbhetcon2} are equivalent to the equations \eqref{hetcon1}-\eqref{hetcon2}, where $\Pi_i$ and $\Gamma_i$ are given by \eqref{gampi}. Moreover, from \eqref{simprob} we know that if $\zeta_i(0) = \mH_T(x_0(0)) \bar \alpha_i$, then $\zeta_i(t) = \mH_T(x_0(t)) \bar \alpha_i$. Notice that $\mH_T(\dot x_0(t)) \bar \alpha_i= A_0 \mH_T(x_0(t)) \bar \alpha_i = A_0 \zeta_i(t)$. Therefore, \eqref{dbdynamicu} is equivalent to \eqref{dynamicu1}. The proof can be straightforwardly completed following similar steps as in the proof of \cite[Theorems~3 and 5]{WielandSepAll2011}.
\end{proof}

In the following section, we test the applicability of the proposed method in a numerical example.

\section{SIMULATIONS}
\label{secsimul}

\begin{figure}
	\begin{center}
		\begin{tikzpicture}[auto,every text node part/.style={align=center}]
			\node[sum] (leader) {$0$};
			\node[midpoint,left of=leader,node distance=2.1cm] (mid1) {};
			\node[sum,below of=mid1,node distance=1cm] (age1) {$1$};
			\node[midpoint,left of=age1,node distance=2cm] (mid2) {};
			\node[sum,below of=mid2,node distance=0.3cm] (age2) {$2$};
			\node[sum,below of=age1,node distance=1.5cm] (age3) {$3$};
			\node[midpoint,right of=age1,node distance=2cm] (mid3) {};
			\node[sum,below of=mid3,node distance=0.8cm] (age4) {$4$};
			\draw[arrow] (leader) -- node {} (age1);
			\draw[arrow] (age1) -- node {} (age2);
			\draw[arrow2] (age1) -- node {} (age3);
			\draw[arrow] (age2) -- node {} (age3);
			\draw[arrow2] (age3) -- node {} (age4);
		\end{tikzpicture}
		\caption{Communication graph for simulation.}
		\label{figgraph}
	\end{center}
\end{figure}
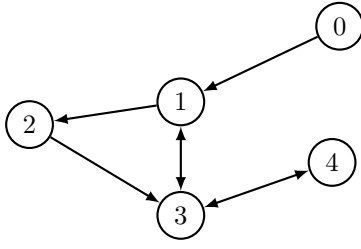

In this section, we apply the results of Theorem \ref{thmhet} to achieve synchronization in a heterogeneous multiagent system\footnote{Matlab code available \url{https://doi.org/10.25835/wj72f6ba}.} \eqref{heteroi}-\eqref{heterol}. We consider one leader and four follower agents connected in the communication graph displayed in Figure \ref{figgraph}. If agent $i$ receives information from agent $j$, then we set $a_{ij}=1$. Since only agent $1$ receives information from the leader, we have $g_1=1$, while $g_i=0$ for $i=2,3,4$. Here we consider the same system models as those used in \cite[Section 4]{WielandSepAll2011}. That is, for the leader \eqref{heterol} we use the matrices
\begin{equation*}
	A_0 = \begin{bmatrix}
		0 & 1 \\ 0 & 0
	\end{bmatrix}, \qquad C_0=\begin{bmatrix}
	1 & 0
\end{bmatrix},
\end{equation*}
while for the follower agents \eqref{heteroi} we use
\begin{equation*}
	A_i = \begin{bmatrix}
		0 & 1 & 0 \\ 0 & 0 & a_i \\ 0 & -b_i & -c_i
	\end{bmatrix}, \quad B_i=\begin{bmatrix}
		0 \\ 0 \\ d_i
	\end{bmatrix}, \quad C_i=\begin{bmatrix}
	1 & 0 & 0
\end{bmatrix},
\end{equation*}
where the constants $\{ a_i,b_i,c_i,d_i \}$ take the values $\{ 1,0,1,1 \}$, $\{ 1,0,10,2 \}$, $\{ 1,10,2,1 \}$ and $\{ 1,1,2,1 \}$, for $i=1,\ldots,4$, respectively. With these models, the agents may correspond to a set of autonomous vehicles with the task of synchronizing their positions along the trajectory of the leader.

\begin{figure}
	\begin{center}
		\includegraphics[height=4.5cm]{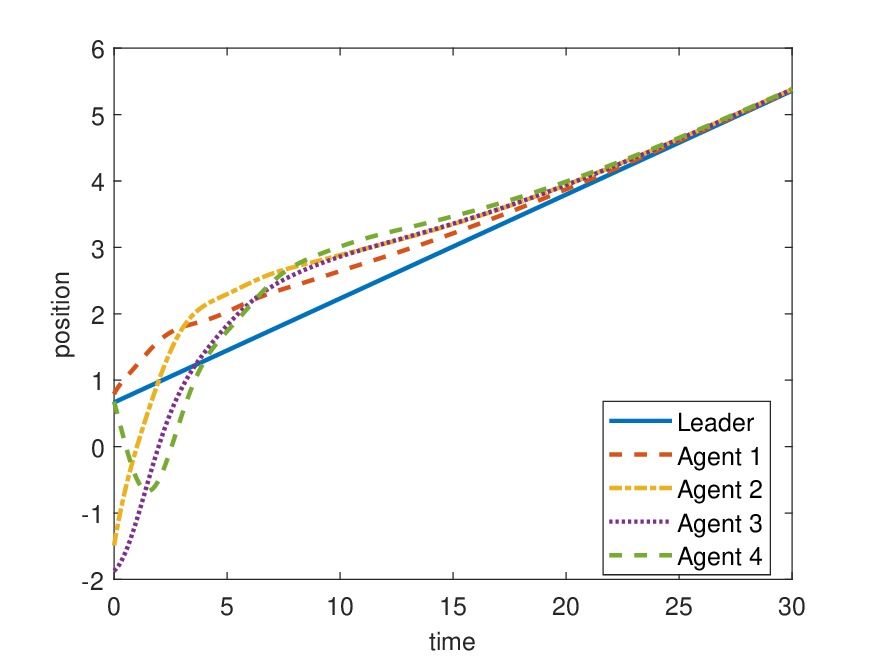}
		\caption{Output trajectories of the leader and the agents.}
		\label{figpos}
	\end{center}
\end{figure}

The procedure described in Theorem \ref{thmhet} is applied. A PCPE input of order $4$ was applied to each follower for data collection. The trajectory of the leader was collected from a random initial condition. The matrices $K_i$ were obtained using the method in \eqref{deper1}-\eqref{deper2}. Finally, the matrices $\Pi_i$ and $\Gamma_i$ were computed and the controllers given by \eqref{dynamicu2} and \eqref{dbdynamicu} were applied to the agents. The resulting output trajectories are shown in Figure \ref{figpos}. It can be observed that synchronization with the leader trajectory was achieved.

\section{CONCLUSIONS}
\label{secconc}

This paper presented novel data-based solutions to the synchronization problem in multiagent systems. First, the problem with identical agents was considered and a solution using LMIs was proposed. This procedure yields distributed static controllers for each agent, where no dynamic variables are required. Then, the output synchronization problem was solved for heterogeneous systems by means of dynamic control inputs without any knowledge of the leader or the followers dynamics. As future work we will investigate conditions to achieve synchronization in heterogeneous agents using LMIs as in \eqref{homsol}, and develop distributed controllers that are robust against noise in the data measurements.


\bibliographystyle{IEEEtran}
\bibliography{IEEEabrv,dbcontrol_refs}


\end{document}